\documentclass[useAMS,usenatbib]{mn2e}
\usepackage{graphicx}     
\usepackage{epsfig}  
\usepackage{times}    
\usepackage{subfigure}
\title[The Hubble constant from galaxy lenses: impacts of triaxiality and model degeneracies]{The Hubble constant from galaxy lenses: impacts of triaxiality and model degeneracies}
\author[Corless, Dobke \& King]{Virginia L. Corless$^{1}$\thanks{E-mail:
vc258@ast.cam.ac.uk}, Benjamin M. Dobke$^{1}$\thanks{E-mail:
bdobke@ast.cam.ac.uk}, Lindsay J. King$^{1}$\thanks{E-mail:ljk@ast.cam.ac.uk}\\
$^{1}$Institute of Astronomy, University of Cambridge, Madingley Road, Cambridge, United Kingdom}
\date{Accepted ... / Received ... }
\pagerange{\pageref{firstpage}--\pageref{lastpage}} \pubyear{2007}

\begin{document}

\label{firstpage}

\maketitle

\begin{abstract}
The Hubble constant can be constrained using the time delays between multiple images of gravitationally lensed sources. In some notable cases, typical lensing analyses assuming isothermal galaxy density profiles produce low values for the Hubble constant, inconsistent with the result of the HST Key Project (72 $\pm$ 8 km\,s$^{-1}$ Mpc$^{-1}$). Possible systematics in the values of the Hubble constant derived from galaxy lensing systems can result from a number of factors, e.g. neglect of environmental effects, assumption of isothermality, or contamination by line-of-sight structures.  One additional potentially important factor is the triaxial structure of the lensing galaxy halo; most lens models account for halo shape simply by perturbing the projected spherical lensing potential, an approximation that is often necessary but that is inadequate at the levels of triaxiality predicted in the CDM paradigm. To quantify the potential error introduced by this assumption in estimates of the Hubble parameter, we strongly lens a distant galaxy through a sample of triaxial softened isothermal halos and use an MCMC method to constrain the lensing halo profile and the Hubble parameter from the resulting multiple image systems.  We explore the major degeneracies between the Hubble parameter and several parameters of the lensing model, finding that without a way to accurately break these degeneracies accurate estimates of the Hubble parameter are not possible. Crucially, we find that triaxiality does {\it not} significantly bias estimates of the Hubble constant, and offer an analytic explanation for this behaviour in the case of isothermal profiles. Neglected triaxial halo shape cannot contribute to the low Hubble constant values derived in a number of galaxy lens systems.
\end{abstract}


\section{Introduction}
Gravitationally lensed, multiply imaged radio source time delays are used to constrain the Hubble constant. Lensing analyses produce a range of values, some of which are significantly lower than the result of the HST Key Project (72 $\pm$ 8 km\,s$^{-1}$ Mpc$^{-1}$); \cite{ogurc} offers an excellent summary of the current state of the field.  Further, although there is a general agreement upon isothermality from a number of differently motivated studies (e.g. \citealt{rix}; \citealt{roma}; \citealt{gerh}; \citealt{treua}; \citealt{koop}; \citealt{sahaa}; \citealt{dobke}), modelling of a number of systems (e.g. PG1115+080, B1600+434, HE2149-2745, and SBS1520+530) has shown that these prefer more centrally concentrated, non-isothermal, density profiles \citep{kocha, kochc}.  The slope of the lensing density profile and the value of Hubble constant are heavily degenerate, and the number of constraints from 
galaxy lens systems small, making it unclear whether the inconsistencies in slope and Hubble constant values indicate flaws in the measurement of both or only one of the two quantities.  

While some recent statistical studies (e.g. \cite{sahaa}) find values consistent with the consensus value, some unexplained inconsistencies remain among the varied results across the field, and whether these stem primarily from external environmental effects, intrinsic physical attributes of the lensing galaxy or universe, or differences in modeling techniques is still unclear. A number of factors are known to influence the derivation of the Hubble constant from galaxy lens systems.  \citet{keet} have shown that lens models of four-image systems overestimate the Hubble constant by up to 15\% when neglecting the effect of the contribution from other group members on the potential of the lens galaxy, with an even higher discrepancy for two-image systems.  Additionally, tidal stripping and shocking has been shown to induce transient fluctuations in the inner density profile of lens galaxies present in groups, and the resultant profile steepening can also contribute to erroneous Hubble constant values if isothermality is assumed \citep{dobkeb}.  Outside the group environment, line-of sight structures can also provide contributions to the lens potential \citep{momc}.  These effects can result in either an over or underestimate of the Hubble constant by virtue of a contributed increase or decrease in the density profile slope.

The intrinsic shape of the lensing galaxy has also been suspected to have a direct bearing on Hubble constant derivation.  \cite{ogurc} recently demonstrated the importance of substructures and external perturbations in galaxy lensing; furthermore, in an investigation with models of 35 galaxy lenses \citet{sahab} proposed that shape-modelling degeneracies (e.g. caused by triaxiality) could also contribute to changes in the time delays and hence the derived Hubble constant values.  Triaxiality is of particular interest because the dark matter halos of $\Lambda$CDM are predicted to be fully triaxial with axis ratios in the mass distribution as low as 0.4
(e.g. \citealt{bett}; \citealt{maccio}), a prediction which observations have loosely confirmed (e.g. \citealt{bak}).  Moreover, it has been shown that the triaxiality of CDM halos can effect the overall lensing probabilities and relative number of different image configurations (double, quadruple, naked cusp lenses) (\citealt{ogurb}; \citealt{rozo}).  
Triaxial studies in galaxy cluster weak lensing have further revealed that neglecting halo shape can result in significant over and underestimates of cluster concentration and mass \citep{corl}.  It seems apparent then that triaxiality can affect a number of critical
observables and derived quantities
in both strong and weak lensing and as such cannot be ignored.  However, due to the small number of 
constraints available in galaxy lensing analyses, halo shape is typically addressed only via an elliptical perturbation to the projected spherical lensing potential, ignoring altogether halo structure along the line-of-sight.  

In this paper we investigate what errors this often necessary neglect of triaxiality in galaxy lensing analyses introduces in 
best fit Hubble parameter values.  We fit an elliptical isothermal mass model to lensing systems of highly (as predicted in simulations) triaxial isothermal galaxy lenses to investigate the maximum discrepancies that can arise from fitting a simplified model to data that in fact originates from a fully triaxial lens.  
We ask:  to what extent can galaxy triaxiality explain the discrepancies between some lensing-derived values for the Hubble constant and those of other methods?
The outline of the paper is as follows:  in \S2 we introduce the softened triaxial isothermal model and its lensing properties, \S3 goes on to present our analysis method and key results, and \S4 draws some conclusions based upon our findings. A standard $\Lambda$CDM cosmology with $H_0 = 72$ km/s/Mpc, $\Omega_m = 0.3$, $\Omega_{\Lambda}=0.7$ is employed throughout.

\section{Lensing by a Softened Triaxial Isothermal Halo} 
To generate a full parameterisation for a softened triaxial isothermal halo we follow the procedure given first in \cite{jing} and implemented for weak lensing in \cite{corl} for a triaxial NFW.  We first generalise the spherical softened isothermal profile to obtain a density profile
\begin{equation}\rho(R) = \frac{\theta_E c_l^2 D_s}{8\pi^2 G D_{ls}\left(S^2 +  R^2\right)}\label{eq:3axrhothetaE}\end{equation}
where $\theta_E$ is an effective Einstein radius, $S$ a triaxial core radius, $R$ a triaxial radius
\begin{equation}R^2 = \frac{X^2}{a^2} + \frac{Y^2}{b^2} + \frac{Z^2}{c^2},\textrm{         }(a\leq b \leq c = 1),\label{eq:3axR}\end{equation}
$a/c$ and $b/c$ the minor:major and intermediate:major axis ratios, respectively, and   
$D_{ls}$ the angular diameter distance between the lens and source and $D_s$ that between the observer and source.

The mass contained within radius $R$ is
\begin{equation}M(<R) = \frac{\theta_E c_l^2 ab D_s}{2 \pi G D_{ls}}\left[R - S \tan^{-1}\left(\frac{R}{S}\right)\right].\label{eq:MinR}\end{equation}
The virial mass is defined as that mass contained within an ellipsoid containing a mean density 200 times the critical density $\rho_c(z)$ at the redshift of the halo, $M_{200} = \frac{800\pi}{3}ab R_{200}^3\rho_c(z)$; combining that definition with (\ref{eq:MinR}) gives an expression for $\theta_E$ as a function of the virial radius $R_{200}$: 
\begin{equation}\theta_E = \frac{1600 \pi^2 \rho_c G D_{ls}R_{200}^3}{3 c_l^2 D_s\left[R_{200} - S \tan^{-1}\left(\frac{R_{200}}{S}\right)\right]}.\end{equation}

\subsection{Lensing Properties}
The full derivation of the lensing properties of a triaxial halo is given by \cite{ogur}, and we summarise some of that work here.  The triaxial halo is projected onto the plane of the sky to find its projected elliptical isodensity contours as a function of the halo's axis ratios and orientation angles ($\theta$, $\phi$) with respect to the the observer's line-of-sight. 
The axis ratio $q$ of the elliptical density contours is given by
\begin{equation}q = \frac{q_Y}{q_X},\label{eq:q}\end{equation}
the elliptical radius by
\begin{equation}\rho^2 = \frac{1}{q_X^2}\left(X^2 + \frac{Y^2}{q^2}\right),\end{equation}
and the orientation angle $\Omega$ on the sky by
\begin{equation}\Omega = \frac{1}{2}\tan ^{-1}\frac{\mathcal{B}}{\mathcal{A}-\mathcal{C}}~~~(q_X \ge q_Y);\label{eq:angle}\end{equation} 
here
\begin{eqnarray}
q_X^2&=&\frac{2f}{\mathcal{A}+\mathcal{C} - \sqrt{(\mathcal{A}-\mathcal{C})^2 + \mathcal{B}^2}}\textrm{ and}\\
q_Y^2&=&\frac{2f}{\mathcal{A}+\mathcal{C} + \sqrt{(\mathcal{A}-\mathcal{C})^2 + \mathcal{B}^2}}, \end{eqnarray}
where
\begin{eqnarray}
\mathcal{A}&=&\cos^2\theta\left(\frac{c^2}{a^2}\sin^2\phi + \frac{c^2}{b^2}\cos^2\phi\right) + \frac{c^2}{a^2}\frac{c^2}{b^2}\sin^2\theta,\\
\mathcal{B}&=&\cos\theta\sin 2\phi\left(\frac{c^2}{a^2} - \frac{c^2}{b^2}\right),\\
\mathcal{C}&=&\frac{c^2}{b^2}\sin^2\phi + \frac{c^2}{a^2}\cos^2\phi, \end{eqnarray}
and
\begin{equation}f = \sin^2\theta\left(\frac{c^2}{a^2}\cos^2\phi + \frac{c^2}{b^2}\sin^2\phi\right) + \cos^2\theta.\label{eq:f}
\end{equation}

General expressions for the lensing potential of a softened elliptical isothermal halo are given in \cite{keeta}; we modify them to incorporate the geometry of the halo projection to establish a direct connection between a 3D halo and its lensing potential.  The lensing potential is
\begin{eqnarray}
\psi(X,Y) &=& \frac{q_x^2}{\sqrt{f}} \bigg\{\mathcal{X}\psi_x + \mathcal{Y}\psi_y  + \theta_E q S \ln\left[\left(1 + q\right)S\right]\nonumber\\
&&- \theta_E q S \ln \sqrt{\left(\Gamma + S\right)^2 + \left(1 - q^2\right)\mathcal{X}^2}\bigg\},\label{eq:psi}\end{eqnarray}
where
\begin{equation}\Gamma^2 = q^2\left(S^2 + \mathcal{X}^2\right) + \mathcal{Y}^2,\end{equation}
\begin{equation}\psi_x = \frac{\theta_E q}{\sqrt{1 - q^2}}\tan^{-1}\left(\frac{\sqrt{1 - q^2}\mathcal{X}}{\Gamma + S}\right),\end{equation}
\begin{equation}\psi_y = \frac{\theta_E q}{\sqrt{1 - q^2}}\tanh^{-1}\left(\frac{\sqrt{1 - q^2}\mathcal{Y}}{\Gamma + q^2 S}\right),\textrm{ and}
\end{equation}
\begin{eqnarray}
\mathcal{X} &=& \frac{X}{q_X};\textrm{    }
 \mathcal{Y} = \frac{Y}{q_X}
 \end{eqnarray}
where $X$, $Y$, and $S$ are all angular distances measured in the image plane.  These analytic formulations were confirmed using the independent numerical methods for calculating lensing properties of elliptical density profiles by integrating over those of their spherical counterparts as given in \cite{keeta} and applied to a triaxial NFW halo in \cite{corl}, as well as by brute force projection of the triaxial halo.

\subsection{Time delays and the Hubble parameter}
The time delays and image positions of strongly lensed systems are completely constrained by the light-travel time equation (e.g. \cite{schneider})
\begin{equation}
t(\vec{ \theta}) = \bigg[\frac{1+z_{l}}{c_l}\bigg]\bigg[\frac{D_{l} D_{s}}{D_{ls}}\bigg]
\bigg[
\frac{1}{2} \left|\vec{ \theta}- \vec{ \beta}\right|^{2} - \psi(\vec{ \theta})\bigg]
\label{eq:time} 
\end{equation}
and its derivative
\begin{equation}
\vec{ \bigtriangledown}t(\vec{ \theta}) = \bigg[\frac{1+z_{l}}{c_l}\bigg]\bigg[\frac{D_{l} D_{s}}{D_{ls}}\bigg]\bigg[(\vec{ \theta} - \vec{ \beta}) - \vec{ \bigtriangledown}\psi\bigg],\label{eq:dtime}\end{equation}
respectively, where $\vec{\theta}$ and $\vec{\beta}$ are angular positions in the lens and source planes respectively, $\psi$ is the lensing potential, and $z_l$ is the redshift of the lens.  
Examining these with some care shows that it is the image positions, and not the time delays between them, that set the normalisation of the model potential: when matching image positions one matches the stationary points of the time-delay surface, thus mandating the model satisfy only the lens equation $\vec{ \theta} - \vec{ \beta} = \vec{ \bigtriangledown}\psi$.  This is observationally independent of the Hubble parameter, for while these are all angular values, with $\vec{ \theta} = \vec{ X}/D_l$, $\vec{ \beta} = \vec{ X}/D_s$, and the physical distances $\vec{ X}$ are not known and cannot therefore be used as a constraint on $D_s$ and $D_l$.  Once the shape and normalisation of the potential is set by the image position constraints via the first derivative of the time delay surface, the normalisation factor $D_lD_s/D_{ls}$ is crucially important in setting the time delays between the multiple images, and it is here the Hubble parameter can be constrained.

\begin{figure}
\epsfig{file=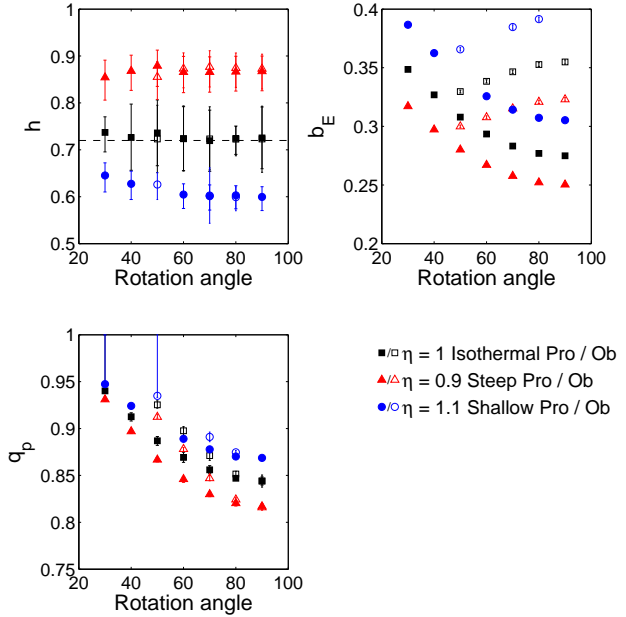,scale=0.45}
\caption{Mean marginalized parameters for oblate and prolate halos with $(a=0.6, b=1.0)$ and $(a=0.6, b=0.6)$ for fixed slope parameters $\eta = \{0.9, 1.0, 1.1\}$ rotated through $90^{\rm o}$ from an orientation of projected circular symmetry at $0^{\rm o}$ to that of maximum visible ellipticity. The dotted line shows the true parameter value of the underlying lens.}
\label{plot:figure1}
\end{figure}
\section{Results \& Discussion}
\subsection{Simulations \& Fitting Method}

\begin{figure*}
\centering
\epsfig{file=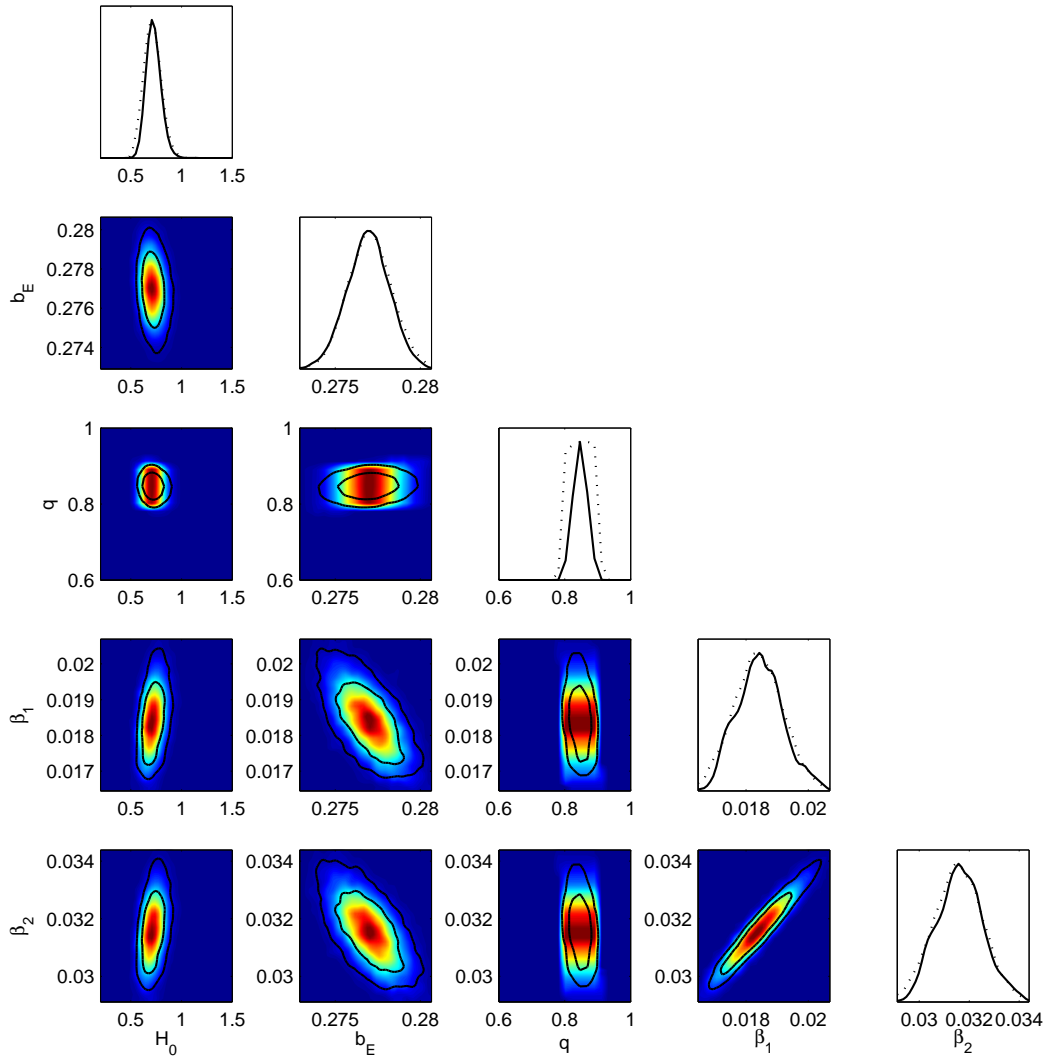,scale=0.8}
\caption{The posterior probability distribution of a power law model with fixed isothermal slope $\eta=1$ fit to an isothermal triaxial prolate halo with axis ratios ($a=0.6$, $b=0.6$) at $80^{\rm o}$, giving elliptical isodensity contours of axis ratio 0.61.  The parameter giving the orientation angle of the isopotential contours $\theta$ is omitted for conciseness, as it shows no degeneracy with any other parameters and is always returned tightly constrained to the true value.  The shading (dotted lines) gives the 2D (1D) likelihood distribution.}
\label{plot:figure2}
\end{figure*}

To generically model the impact of unaccounted for triaxial structure in isothermal galaxies we use a Markov Chain Monte Carlo (MCMC) sampling method to explore the posterior probability distribution obtained when fitting an elliptical power law model to a triaxial isothermal lens.  First, simulated multiple-image lens systems are generated by lensing through a triaxial isothermal halo according to Eq. \ref{eq:psi} and the lens equation.  These simulated observations of image positions and time delays are then fit using a general 2D elliptical power law model governed by seven free parameters.  Such models cannot be constrained using two-image multiple-image systems, as these provide only 5 constraints (2 image positions vectors + 1 time delay); 4-image systems fare better, providing 11 constraints (4 image position vectors + 3 time delays).  However, even in 4-image galaxy lens systems, the data is typically able to meaningfully constrain only one of either the slope of the underlying galaxy potential or the Hubble parameter, due to an oft-noted strong degeneracy between these two key parameters.

Several papers have explored constraints on $H_{\rm o}$ from multiple image time delays, and degeneracies with parameters describing the lens model, in particular the radial profile. Early work includes \cite{goren} and \cite{refsdal}. Later, \cite{witt} consider analytic time delays and estimates of $H_{\rm o}$ from a general family of lensing potentials which allows for angular structure. Their treatment was extended and generalised by \cite{wuck}. \cite{kocha} proposes a semianalytic model to understand observed time delays, isolating the important physical properties of the lens models. For example, building on the treatment of \cite{goren}, he obtains scaling solutions for $H_{\rm o}$ for general lenses which depend on the mean convergence inside annuli bounded by the multiple images.   Due to the strong degeneracy, in most of these and other analyses some assumption is often made regarding the shape of the galaxy density profile and the Hubble parameter then obtained under that assumption (though simultaneous fits have been succesfully carried out in statistical studies, e.g. \cite{sahaa}).  Most often the galaxy halo is taken to be isothermal with density profile slope $\rho(r) \propto r^{-2}$, a well-motivated simplification derived from previous studies of galaxy halo profiles (see eg. \cite{koop}).  However, for any one galaxy, it is quite unlikely that the slope of the profile will be exactly isothermal.  Given the strong degeneracy between the Hubble parameter and the density profile slope, such systematic errors may be quite important.

To avoid having to make such a potentially error-inducing assumption, we adopt an MCMC fitting method that is capable of exploring the full posterior probability distribution and all its degeneracies.  MCMC methods employ a ``guided'' random walk that returns a sample of points representative of the posterior probability distribution; the probability of a certain region of parameter space containing the true model is directly proportional to the density of points sampled in that region.  From the distribution of sample points the full posterior probability distribution is obtained, which is easily and direcly marginalized over to obtain fully marginalized most-probable parameter estimates for all parameters.  These most-probable parameter values reflect the full shape of the posterior probability distribution without assumption about the error distribution, and with full knowledge of the prior.  Such methods have exploded in popularity recently, and there are several excellent references describing the method in detail, e.g. \cite{lewis}, \cite{mackay}.  Simply put, the sampler functions by stepping through parameter space by taking random steps governed by a transfer function, usually a simple $n$-D Gaussian, where $n$ is the number of parameters of the fitted model.  If the randomly-chosen next step is to a point of higher probability than the current position, the step is taken.  If the next step is to a point of lower probability, the step is taken with probability $p({\rm new})/p({\rm current})$.  Thus, the MCMC sampler spends most of its time in high probability regions, but can move ``downhill'' to regions of lower probability in order to explore the entire space and sample all regions of high probability.  Crucially, this method is able to return a true representation of the full posterior probability distribution, regardless of the complexities of that probability distribution, e.g., tight degeneracies, multiple islands of high probability, or a very flat distribution due to poor constraints from the data.

Thus, we are able to fully understand the impacts of fitting 2D models to systems with 3D triaxial structure, without inducing errors by making other limiting assumptions.  To implement the method, we must define the posterior probability function, defined in Bayesian statistics as
\begin{equation}p(\pi|\theta) = \frac{p(\theta|\pi)p(\pi)}{p(\theta)}\end{equation}
where $p(\theta|\pi)$ is the likelihood of the data given the model parameters (the standard likelihood), $p(\pi)$ is the prior probability distribution for the model parameters (e.g. an outside constraint on the slope of the density profile), and $p(\theta)$ is a normalising factor called the {\it evidence}, of great value in comparing models of different classes and parameter types, but expensive to calculate and unnecessary for the accurate exploration of the posterior distribution.  The assignment of priors is often a controversial exercise, but in this case it is straightforward to assign weak, physically sensible priors that do not significantly affect our results.  We define the likelihood function as a standard exponential function of the differences between the model image positions and time delays and those of the ``observed'' data:
\begin{eqnarray} \mathcal{L} &=& \Pi_{i=1}^{4}\exp\left[-\frac{\left({\bf \theta_i} - {\bf \theta_{M,i}}\right)^2}{2\sigma_x^2}\right]\nonumber\\
&\times& \Pi_{i=1}^3\exp\left[-\frac{\left(\Delta t_i - \Delta t_{M,i}\right)^2}{2\sigma_t^2}\right],\end{eqnarray}
and take the measurement errors in the image positions $\sigma_x$ and time delays $\sigma_t$ to respectively be $0.7\%$ and $14\%$, reasonable values for strong lensing analyses.

In our MCMC sampler we employ a 7D two-sided Gaussian transfer function, and use the covariance matrix of an early run to sample in an optimised basis aligned with the major degeneracies of the posterior (of which there are several!).  We tune the step sizes of the sampler to achieve an average acceptance rate of 1/3 in each basis direction, run three independent MCMC chains, started at randomly chosen positions in parameter space, for each lens system, and sample the distribution space until the standard var(chain mean)/mean(chain var) indicator is less than 0.2, indicating chain convergence.  We utilise the GetDist package from the standard CosmoMC (\citealt{lewis}) distribution to calculate convergence statistics, parameter contours, and marginalized parameter estimates.  The highly probable region of parameter space is of very small volume compared to the that of the prior parameter space, and we find it is a significant time-saver to first follow an MCMC chain with very large step sizes to find regions of high probability, then restart a chain at this randomly found, but advantageously chosen, starting point with a step size tuned to the size and shape of the posterior distribution.  This does not upset the statistics of the MCMC method; it is equivalent to throwing away points from the very long burn-in phase that would be required if the step size were kept tuned to the small steps best for exploring the high probability region.

\begin{figure}
\epsfig{file=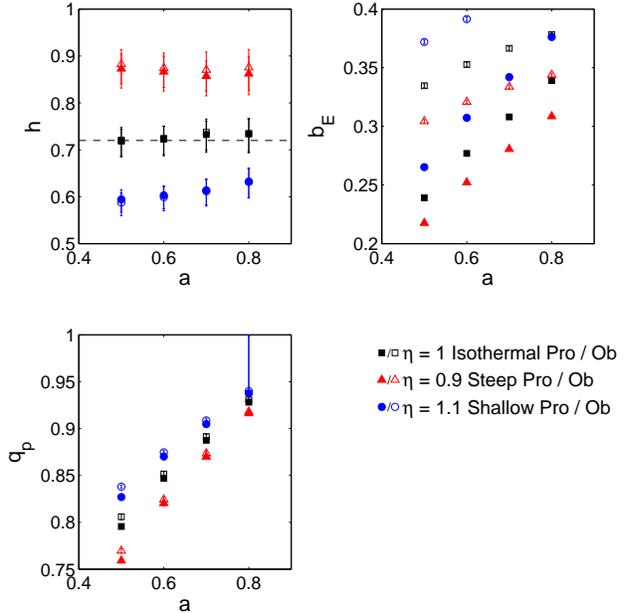,scale=0.45}
\caption{Mean marginalized parameters for oblate and prolate halos with a range of triaxial minor axis ratios $a=\{0.5, 0.6, 0.7, 0.8\}$ at a rotation angle of $80^{\rm o}$ for fixed slope parameters $\eta = \{0.9, 1.0, 1.1\}$.  The dotted lines show the true parameter values of the underlying lens.}
\label{plot:figure3}
\end{figure}

In every case we fit a power-law lensing density profile model to the simulated lensed images and time delays, given by \cite{barkana} (here written with a factor of $2$ modification to match a standard elliptical isothermal model when $\eta=1$), with lensing potential
\begin{equation} \psi(\vec{ \theta}) = \frac{b_{ E}^2}{\eta^2}\left[\frac{X^2 + Y^2/q_p^2}{b_{ E}^2}\right]^{\eta/2},\end{equation}
where $q_p$ is the axis ratio of the isopotential ellipses.  Note that here the ellipticity is in the 2D potential, as opposed to our triaxial model in which it is introduced in the 3D isodensity surfaces.  At high ellipticities ($\epsilon_p > 0.3$) where the projected axis ratio is small ($q_p < 0.54$), near-isothermal elliptical lensing potential models break down, giving unphysical ``dumbbell''-shaped projected mass distributions (\cite{kassiola}).  However, in the regime of ellipticities we study here ($q_p > 0.75$), this problem is insignificant, and we choose this elliptical potential model as representative of those most often fit in lensing analyses and for its analytical simplicity.   The potential is always less elliptical than the underlying mass distribution, such that the axis ratio $q_p$ is always greater than the axis ratio $q$ of the isodensity contours. The underlying density profile steepens with decreasing $\eta$, such that $\eta=0$ corresponds to a Hubble density profile that goes as $r^{-3}$.  Thus, the seven free parameters are the profile slope parameter $\eta$, the effective Einstein radius $b_{E}$, the axis ratio of the model isopotential contours $q_p$, the orientation angle of those contours $\theta$, the source position components $\beta_1$ and $\beta_2$, and the Hubble parameter $h$.

We begin our analysis with a simple case, similar in method to many typical lensing analyses, in which we set the slope parameter to a range of fixed values and carry out parameter fits, and examine the impacts of triaxiality in that simplified framework.  We then offer an analytic explanation for our numerical results.  Finally, we examine the problem using the fully generalised power-law model in which the slope is allowed to vary freely, to complete our understanding of the role of triaxiality and to better understand the larger degeneracies that govern the problem of Hubble constant determination from strong lensing systems.
\begin{figure*}
\centering
\epsfig{file=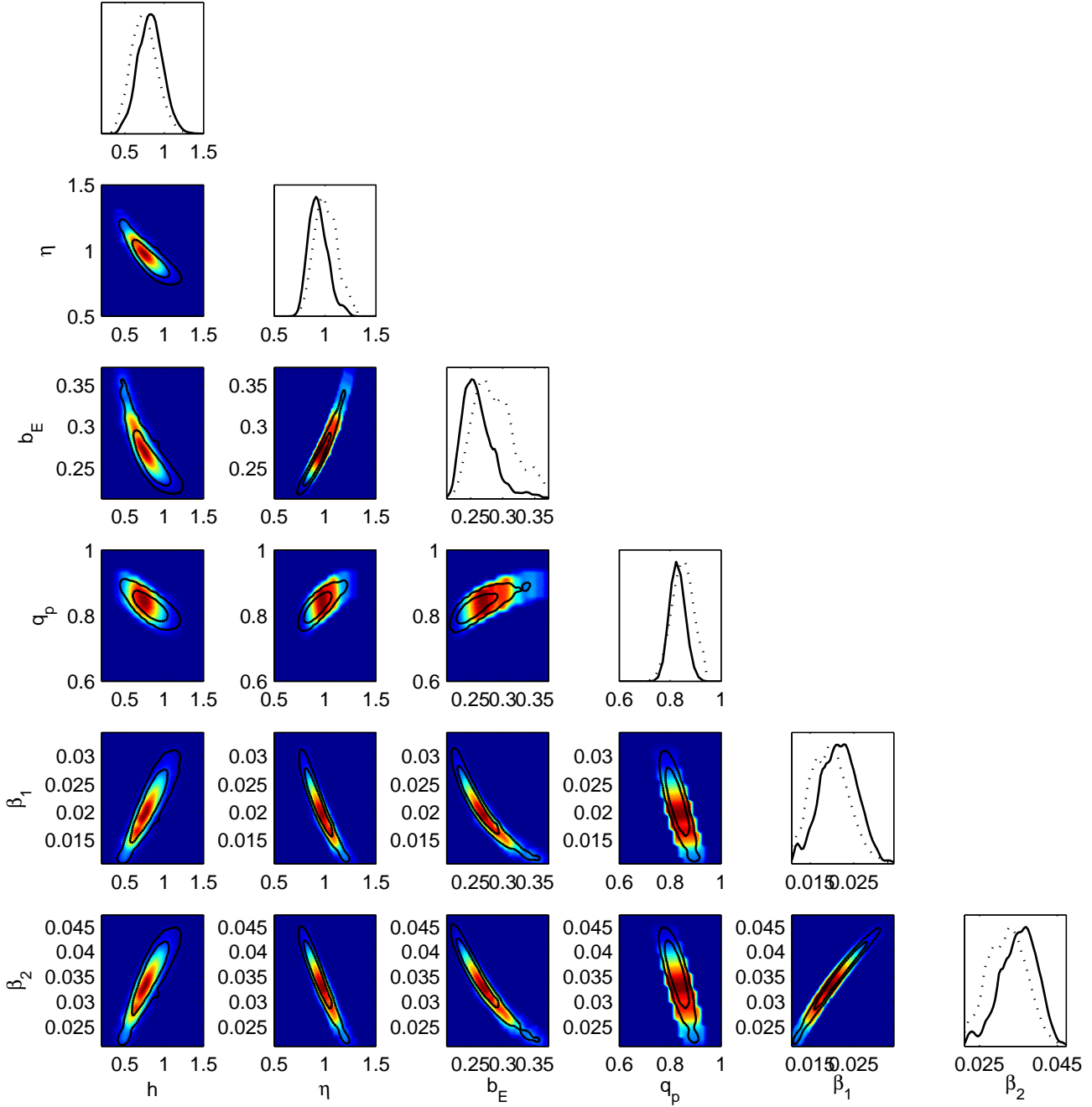,scale=1.0}
\caption{The posterior probability distribution of a general power law model fit to an isothermal triaxial prolate halo with axis ratios ($a=0.6$, $b=0.6$) oriented at $80^{\rm o}$, giving elliptical isodensity contours of axis ratio 0.61.  The parameter giving the orientation angle of the isopotential contours $\theta$ is omitted for conciseness, as it shows no degeneracy with any other parameters and is always returned tightly constrained to the true value.  The shading (dotted lines) gives the 2D (1D) likelihood distribution.}
\label{plot:figure4}
\end{figure*}
\begin{figure*}
\centering
\epsfig{file=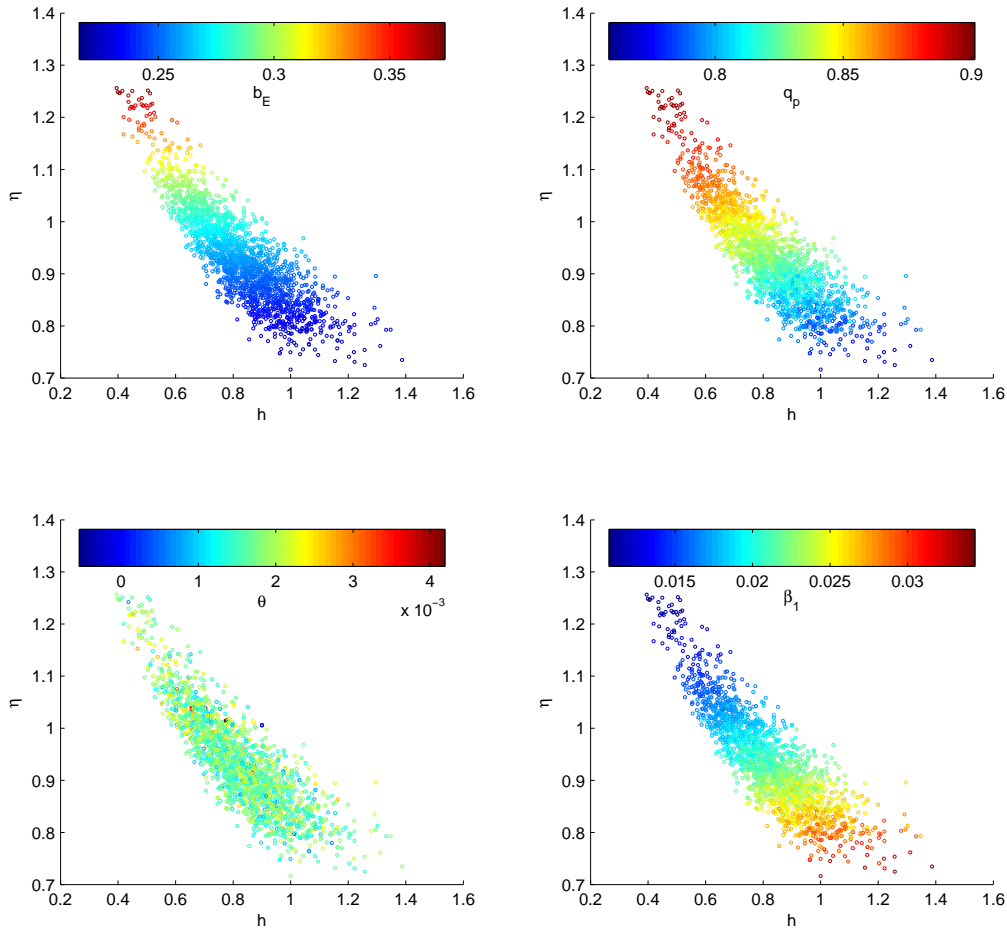,scale=0.7}
\caption{Projections of the posterior probability distribution into the $\eta$-$h$ plane of a power law model fit to an isothermal triaxial prolate halo with axis ratios ($a=0.6$, $b=0.6$) oriented at $80^{\rm o}$, giving elliptical isodensity contours of axis ratio 0.61 , shaded by the other important model parameters, respectively, $b_E$, $q_p$, $\theta$, and $\beta_1$.  Note the strong interdependence of the $\eta$-$h$ degeneracy on $b_E$, $q_p$, and ${\bf \beta}$.}
\label{plot:figure5}
\end{figure*}


\subsection{A Fixed-slope model fit to oblate and prolate halos}\label{sec:FS}
To begin to isolate the impacts of triaxiality in the problem of Hubble parameter determination from strong galaxy lensing systems, and to allow for direct comparisons to standard methods employed in many analyses (e.g. \cite{kochb}), we fix the density profile slope at one of three values and carry out an MCMC exploration of parameter space.  We fix the slope of the underlying halo to either isothermal ($\eta = 1$), steeper than isothermal ($\eta = 0.9$), or shallower than isothermal ($\eta = 1.1$), approximating the errors between the true and model slopes sure to sometimes occur if the slope is externally constrained to avoid the strong degeneracy between it and the Hubble parameter.  We then fit these models to lens systems of symmetric oblate and prolate triaxial lensing halos at $z=0.5$ with a 1pc core with axis ratios ($a=0.6, b=1.0$) and ($a=0.6$, $b=0.6$), respectively, each containing a mass of $1\times10^{11}$ M$_{\odot}$ inside the triaxial effective Einstein radius $R_E = X_E^2/a^2 + Y_E^2/b^2 + Z_E^2 = \theta_ED_l$.  

\subsubsection{Hubble parameter as a function of triaxial orientation}\label{sec:FS1}
We fit to each halo type at ten different orientations stepped from $0^{\textrm o}$ to $90^{\textrm o}$; at $90^{\textrm o}$ each halo looks maximally elliptical with projected isodensity axis ratio $a$ while at $0^{\textrm o}$ the projected profile is circular.  When the projected ellipticity is small only 2 images are formed, and we discard these systems because they provide too few constraints to meaningfully study the Hubble parameter, leaving 7 prolate systems and 5 oblate systems for study (there are more 4-image prolate systems because prolate halos are more efficient lenses than oblate, such that their threshold for 4-image systems is at a lower ellipticity: see \cite{corl}).  At each orientation the halos lens a source at $z=2$ positioned inside the multiple-imaging region at $\vec{ \beta} = \{0.02'', 0.03''\}$.  The image positions are found via the lens equation using a standard rootfinder, and the time delays between images calculated via the time delay surface (Eq. \ref{eq:time}); central images are discarded due to their high level of demagnification, which makes them invisible in observations (e.g. \cite{wall}).

The results are plotted in Figure~\ref{plot:figure1}.  Crucially, when the slope of the fitted halo matches that of the underlying lens -- when they are both isothermal -- the Hubble parameter is accurately returned in every case!  Thus, when the slopes are accurately matched in lens modelling, triaxiality induces {\it no error} in estimates of the Hubble parameter.  The triaxial cases follow the familiar trends observed when there is a mismatch between the model slope and the true slope of the lens: as e.g. \cite{kocha} has previously shown, fitting a model shallower than the true potential leads to underestimates of the Hubble parameter, while overestimates are apparent when the fitted model has a steeper slope than the true lensing halo.  The error in the returned Hubble parameter is worst when the visible axis ratio is lowest and the lensing potential most elliptical.

Figure \ref{plot:figure2} shows the posterior probability distribution when the power law model is fit to a prolate halo oriented at $80^{\rm o}$ with the slope set to match the true value of the isothermal lensing halo $\eta=1$.  The orientation angle of the isopotential contours $\theta$ is omitted for conciseness; it shows no degeneracies with any other parameter and is very tightly constrained to the true value.  Importantly, the only significant degeneracy visible is between the the components of the source position. Thus, when the slope is constrained accurately, a meaningful and accurate value for the Hubble parameter can be derived from strong lens systems!  However, the slope must indeed be {\it accurately} constrained; Figure \ref{plot:figure1} shows clearly that a mismatch between the true slope value and the model slope value yields signficantly incorrect values for the Hubble parameter.

\subsubsection{Hubble parameter as a function of triaxial axis ratio}
To complete this portion of the analysis, we seek to further clarify any impact of triaxiality by calculating the best-fitting parameter values for oblate and prolate halos with a range of triaxial minor axis ratios $a=\{0.5, 0.6, 0.7, 0.8\}$, all oriented at $80^{\rm o}$ so as to have high projected ellipticities, at the same three fixed values of $\eta$.  The results are plotted in Figure \ref{plot:figure3}.  Again, when the density profile slope is mismatched to the true underlying value, increased projected ellipticity exacerbates the error induced in the Hubble parameter, with the best-fit value of $h$ approaching slightly closer to the true value as the minor axis ratio $a$ moves closer to 1.  That this trend occurs both for halos of the same fundamental shape oriented at different angles and for halo of different shape oriented at the same angle, and for both prolate and oblate halos, indicates it is purely associated with the visible ellipticity, with no dependence on the (very different) 3D structure of the various halos.

We return to the point that there is essentially no difference in the value of the recovered Hubble parameter if the halo is oblate or prolate.  As noted, the geometry of these two cases is quite different.  When oriented at $80^{\rm o}$ a prolate halo is like a rugby ball being looked at edge on -- its long axis is visible in the plane of the sky and there is relatively little mass along the line of sight.  Conversely, the oblate halo is like a pancake looked at edge on, with its one short axis in the plain of the sky and a large amount of mass along the line of sight. This difference in projected mass is reflected in the potential normalisation $b_E$.  However, the Hubble parameter is in no way effected by this significant difference in 3D structure!  The line-of-sight structure of the lensing halo appears unimportant to the Hubble parameter problem.

\begin{figure}
\epsfig{file=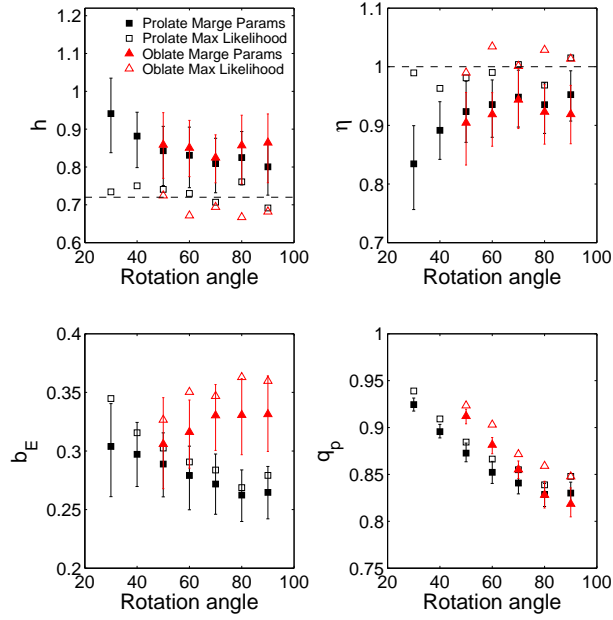,scale=0.45}
\caption{Marginalized mean best-fitting parameters for symmetric oblate and prolate halos of minor axis ratio $a=0.6$ (filled symbols), as well as the maximum-likelihood parameters (unfilled), plotted as a function of the rotation angle of the lensing halo: at $0^{\rm o}$ the halos are circularly symmetric in projection, at $90^{\rm o}$ they show a maximum visible ellipticity of 0.6 in the projected isodensity contours.  The dotted lines show the true parameter values of the underlying lensing halo.}
\label{plot:figure6}
\end{figure}

\subsection{An analytic evaluation of the impact of triaxiality on $H_{\rm 0}$}
To better understand this finding that Hubble parameter estimation is unaffected by the line-of-sight structure of triaxial halos, we look to the case of triaxial isothermal halos fitted with elliptical isothermal models, and more closely examine the expressions for the triaxial isothermal potential and its gradient, and their role in determing H$_0$ in a strong lensing system. The potential of the softened triaxial isothermal model was given in Eq.\ref{eq:psi}; its derivatives are

\begin{eqnarray}\frac{\partial \psi}{\partial \theta_X}&=&\frac{\theta_Eqq_x}{\sqrt{f}}\bigg\{\frac{1}{\sqrt{1-q^2}}\tan^{-1}\left(\frac{\sqrt{1-q^2}\mathcal{X}}{\Gamma + S}\right)\nonumber\\
&+&\frac{\mathcal{X}}{\left(\Gamma+S\right)^2+\left(1-q^2\right)\mathcal{X}^2}\left[\Gamma - \frac{q^2}{\Gamma}\left(\mathcal{X}^2 + S^2\right)\right]\nonumber \\
&-&\frac{\mathcal{Y}^2\mathcal{X}q^2}{\Gamma\left[\left(\Gamma+q^2S\right)^2 - \left(1-q^2\right)\mathcal{Y}^2\right]}\bigg\},\label{eq:dpsi1}\\
\frac{\partial \psi}{\partial \theta_Y}&=&\frac{\theta_Eqq_x}{\sqrt{f}}\bigg\{\frac{1}{\sqrt{1-q^2}}\tanh^{-1}\left(\frac{\sqrt{1-q^2}\mathcal{Y}}{\Gamma + q^2S}\right)\nonumber \\
&+&\frac{\mathcal{Y}(\Gamma+q^2S)}{\left(\Gamma+q^2S\right)^2 - \left(1-q^2\right)\mathcal{Y}^2}\left(1 - \frac{\mathcal{Y}^2}{\Gamma\left(\Gamma+q^2S\right)}\right)\nonumber\\
&-&\frac{\mathcal{Y}\left(\mathcal{X}^2 + S\Gamma+S^2\right)}{\Gamma\left[\left(\Gamma+S\right)^2 + \left(1-q^2\right)\mathcal{X}^2\right]}\bigg\}.\label{eq:dpsi}
 \end{eqnarray}

Examining Eqs. \ref{eq:psi}, \ref{eq:dpsi1}, and \ref{eq:dpsi} we see that, when $S$ is small, the triaxial structure of the halo manifests itself only through multiplicative prefactors.  This is not a generic behaviour for all potentials; it is true of this model because, when $S$ is small, all terms carry factors of the same power of $q_X$, the normalising ratio that carries information about the length of the axis hidden along the line of sight. All terms in the potential that are important when $S$ is small are proportional to $\mathcal{X}$ or $\mathcal{Y}$, giving each term an overall normalisation of $q_X$, and similarly all important terms in its derivative are dimensionless in lengths giving each term a normalisation of $q_X$.   When $S$ is large, or were there any terms that carried unequal powers of $q_X$, this simple multiplicative renormalisation due to triaxial structure would no longer hold.  However, given that there is good evidence that galactic cores are very small (\cite{wall}) and that galaxy density profiles are generally isothermal over the central region to which strong lensing is sensitive, this straightforward analysis indicates that triaxiality will not independently lead to any significant change in the shape of the lensing potential and thus creates no significant errors in the measurement of the Hubble parameter, as indicated in our simulations.  The multiplicative factors introduced by the triaxial structure of the lensing halo are apparent in the varied values of the normalisation of the potential $b_{\rm E}$ returned for different orientations of the triaxial halos.

\subsection{A general model fit to symmetric oblate and prolate halos}
To complete our understanding of the problem of Hubble parameter estimation from triaxial lens systems, we now look to the fully general case in which the elliptical power law model is fit with both Hubble and slope parameters left free to vary. 
 
\subsubsection{Hubble parameter as a function of triaxial orientation: most-probable parameters}
We fit this general model to the same family of halos described in Section \ref{sec:FS1}.  The complete posterior probability distribution for one of the halos, the prolate halo oriented at $80^{\rm o}$ such that the ellipticity is approaching the maximum, is shown in Figure \ref{plot:figure4}, again absent the well-constrained and independent orientation angle $\theta$ to make the plot clearer. The well-known degeneracy between the Hubble parameter $h$ and the slope parameter $\eta$ is clearly visible, but interestingly, is not the only strong degeneracy present.  Both the Hubble parameter and the density profile slope are also strongly degenerate with the normalisation of the potential $b_E$, and with the source position {\bf $\beta$}.

The regions of highest likelihood (indicated by the shading) are centred on the true parameter values; however, the most-probable parameter values obtained by marginalizing over all other parameters give consistently high values for the Hubble parameter and low values for the profile slope, indicating that while the true parameter values give a slightly higher likelihood value than others, there are other likely parameter value combinations including low values for $\eta$ and high values for $h$ that occupy significantly larger volumes in parameter space.  Were the errors in this problem Gaussian, the maximum-likelihood parameter estimates should correspond to the most-probable marginalized parameter value; however, as seen in Figure \ref{plot:figure4}, the posterior probability distribution is highly non-Gaussian, leading to the separation of the point of maximum likelihood from the point of maximum probability.   The marginalized most-probable parameter values represent the true 'best' model; thus, left unbroken, the degeneracies of the problem produce real errors in any lensing result.

To better picture the way these degeneracies function, Figure \ref{plot:figure5} plots the posterior probability distribution in the $\eta$-$h$ plane, coloured by, respectively, each point's $b_E$, $q_p$, $\theta$, and $\beta_1$ value.  These clearly show that the degeneracy between the slope and Hubble parameter is tightly connected to the normalisation of the potential $b_E$, the axis ratio $q_p$ of the isopotential contours, and the source position {\bf $\beta$}, but is completely independent of the orientation angle $\theta$.  Thus, we should not speak just of a $\eta$-$h$ degeneracy, but rather of a strong $\eta$-$h$-$b_E$-$q_p$-${\bf \beta}$ degeneracy!  This demonstrates that 4-image lens systems cannot meaningfully constrain the Hubble parameter with significance unless something can be done to at least partially break this degeneracy. Constraints on the profile slope or normalisation seem the best hope, as it impossible to learn anything of the source position by methods other than lensing.  The shape of this degeneracy now makes clear the source of the error in Hubble parameter and slope returned when the slope is fixed to the wrong value, as in Figure \ref{plot:figure1}; setting the slope is equivalent to drawing a line at a particular value of $\eta$ in the scatter plots of Figure \ref{plot:figure5} and taking only the points that fall on that line.  If the line is taken at the wrong place, completely adequate models will be found that will return completely false values for the other model parameters.

\begin{figure}
\epsfig{file=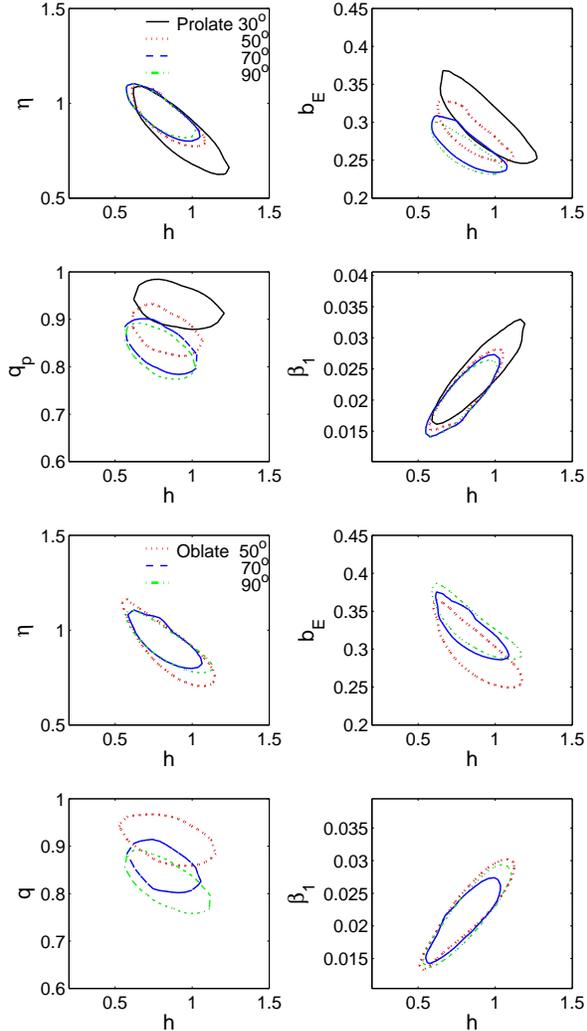,scale=0.45}
\caption{1-$\sigma$ (68$\%$) confidence contours for power law model parameters $\eta$, $b_E$, $q_p$, and $\beta_1$ with Hubble parameter for a sampling of symmetric oblate and prolate halos of triaxial minor axis ratio $a=0.6$ at a range of rotation angles and thus projected ellipticities.}
\label{plot:figure8}
\end{figure}
Figure \ref{plot:figure6} plots the marginalized mean best-fitting and the maximum likelihood values for the Hubble parameter $h$, the profile slope parameter $\eta$, the potential normalisation $b_E$, and the axis ratio of the isopotential ellipses $q_p$ obtained for the prolate and oblate halos as a function of orientation angle.  We find that the most-probable values of $h$ and $\eta$ vary inversely, with $h$ getting smaller and $\eta$ larger with increasing visible ellipticity.  The potential normalisation $b_E$ decreases significantly for the prolate halos and increases significantly for the oblate halos with increasing ellipticity; this is as expected, because in the maximum ellipticity orientation prolate halos have minimal mass hidden along the line-of-sight, and oblate halos maximum, and the converse in the circularly-symmetric orientation at $0^{\rm o}$.  This result is consistent with the behaviour found in previous analyses that weak lensing mass estimates are strongly effected by unaccounted-for triaxiality \citep{corl}.  The values of the axis ratio of the isopotential contours $q_p$ exhibit the trend expected, decreasing as the visible ellipticity increases from orientation $0^{\rm o}$ to $90^{\rm o}$.

\subsubsection{Hubble parameter as a function of triaxial orientation: posterior probability distributions}
To better understand what is happening to the posterior probability distribution as the triaxial halo orientation is changed, in Figure \ref{plot:figure8} we plot the 1-$\sigma$ (68$\%$) confidence contours for $\eta$, $b_E$, $q_p$, and $\beta_1$ against $h$ for prolate and oblate halos at several orientations, again with visible ellipticity increasing with rotation angle.  For both oblate and prolate halos, the $\eta$-$h$ contours maintain approximately the same position, but are tighter at higher visible ellipticities.  The least elliptical case alos shows an extention of its contour to lower values of $\eta$ and higher values of $h$.  The contours for the potential normalisation $b_E$-$h$ move down to lower normalisations for the more elliptical prolate halos, and up to higher normalisations for more elliptical oblate halos, as expected given the line-of-sight masses of halos of these geometries.  Further, they decrease in size as ellipticity increases, as with the $\eta$-$h$ contours.  For both prolate and oblate halos the contours for isopotential axis ratio $q_p$-$h$ move down to lower values as the ellipticity increases, as expected.  The $\beta_1$-$h$ contours behave very similarly to those for $h$-$\eta$, maintaining similar positions in parameter space but changing size.

To understand the positioning and size of these contours we consider the nature of the multi-variable degeneracy.  It lies in parameter space such that high values of the axis ratio $q_p$ correspond to high values of the slope parameter $\eta$, and low values for the Hubble parameter $h$ and source position ${\bf \beta}$.  Now, for halos with low amounts of visible ellipticity and thus true values of $q_p$ near 1, there is less space in parameter space for models at values of $q_p$ higher than the true value, because the posterior distribution is truncated with a hard wall of zero probability at $q_p=1$.  Thus, the contours will extend proportionally more into the regions of lower than true $q_p$, corresponding to lower values of $\eta$ and higher values of the Hubble parameter $h$ and source position ${\bf \beta}$. This trend is exactly what is seen as we move from the low-$q_p$ halos (rotation angles near $90^{\rm o}$) to the high-$q_p$ halos aligned closer to the circularly symmetric ($0^{\rm o}$) position.

In addition to their position, the high-$q_p$ halos also produce larger 1$\sigma$ contours.  This change in contour size cannot be due to the lensing strength: in strong lensing, so long as the same number of images are resolved, the relative strength of the lens and the scale of the image separations does not change the amount of information available to constrain the lens system.  This is born out in the results, in that it is the high-$q$ models in both oblate and prolate cases that exhibit larger contours; in the prolate case these are indeed the stronger lenses with higher projected masses, but in the oblate case they are the weaker lenses, with very low lensing convergences.  Instead, this again appears to be the natural product of the truncation of the posterior probability distribution at $q_p=1$, which shifts and extends the 1$\sigma$ region towards lower values of $q_p$ and $\eta$ and to higher values of $h$.  This effect -- not an error or a problem but a true representation of the shape of the posterior probability distribution -- is illustrated and explored in more detail in Appendix \ref{sec:appa}.

\subsubsection{Hubble parameter as a function of triaxial axis ratio}
\begin{figure}
\epsfig{file=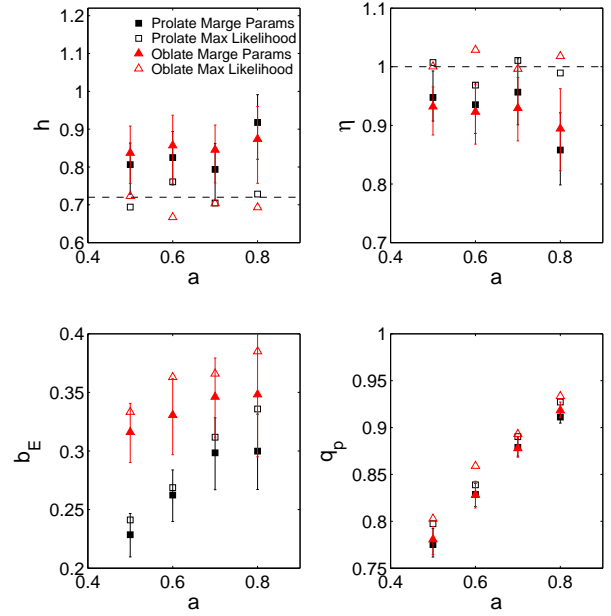,scale=0.45}
\caption{Marginalized mean best-fit parameters $h$, $\eta$, $b_E$, and $q_p$ for symmetric oblate and prolate lensing halos of minor axis ratio $a=\{0.5, 0.6, 0.7, 0.8\}$ (filled symbols) at a rotation angle of $80^{\rm o}$, as well as the maximum likelihood parameters (unfilled), plotted as a function of the minor axis ratio.  The dotted lines show the true parameter values of the underlying lensing halo.}
\label{plot:figure9}
\end{figure}

Again, to understand the importance of the degree of triaxiality, now in the general case, we carry out the same analysis as above but for prolate and oblate halos of axis ratios $a=\{0.5, 0.6, 0.7, 0.8\}$, all oriented at $80^{\rm o}$.  Figure \ref{plot:figure9} plots the marginalized best-fit parameters $h$, $\eta$, $b_E$, and $q_p$ as a function of minor triaxial axis ratio $a$ for both prolate and oblate halos.  Again, increasing the amount of triaxiality does {\it not} significantly change the best-fitting slope or Hubble parameter within the errors.  There is a weak trend toward there being larger offsets from the true value at {\it smaller} visible ellipticities, the opposite of that seen in the fixed-slope case.  This trend can be understood as an extension of that seen for this general case as a function of orientation angle; again, the 1$\sigma$ region is shifted towards lower values of $q_p$ and $\eta$ and to higher values of $h$ for lower visible ellipticity cases; again see Appendix \ref{sec:appa} for further discussion.
 
\begin{figure}
\epsfig{file=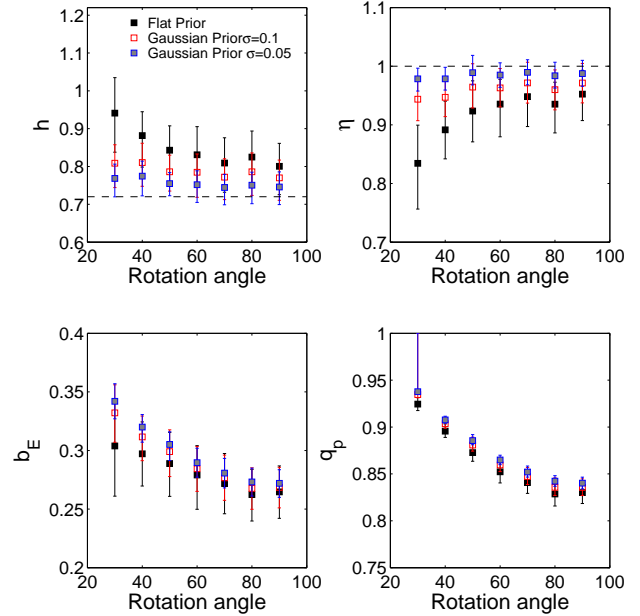,scale=0.45}
\caption{Marginalized mean best-fitting parameters for symmetric prolate halos of minor axis ratio $a=0.6$ under three different priors on the density profile slope $\eta$, plotted as a function of the rotation angle of the lensing halo: at $0^{\rm o}$ the halos are circularly symmetric in projection, at $90^{\rm o}$ they show a maximum visible ellipticity of 0.6 in the projected isodensity contours.  The dotted lines show the true parameter values of the underlying lensing halo.  A tighter prior centred around the true slope value moves all the marginalized parameter values towards the true value.  Such a prior may often be justified by independent knowledge of the galaxy density profile.}
\label{plot:figure7}
\end{figure}

\subsubsection{Priors and best-fitting parameters}
These results represent those of the most general analysis, taking wide, flat priors on all parameters ($0.2 < h< 1.5$; $0.5<\eta <1.5$; $0.1<q<1.0$; $b_E \ge 0$; {\bf $\beta$} and $\theta$ unconstrained).  However, in most cases, it is possible and indeed preferable to impose a physically realistic prior on the density profile slope, as work on galaxy profiles converges to show that nearly all galaxies are very close to isothermal.  To understand the impact on our results of including this physically motivated prior we impose two different Gaussian priors on the slope parameter $\eta$, centred on the isothermal value $\eta=1$, with $\sigma = 0.1$ and $\sigma=0.05$. The marginalized best-fit parameters found under these prior are shown in Figure \ref{plot:figure7}.  As expected, the best-fitting parameters migrate towards the true values of the system when the prior is tightened on the slope, as the $\eta -h$ degeneracy is partially broken in favour of the true values of the lensing model employed in the simulations.

We note a significant offset from the true values in the most-probable marginalized parameters under a flat prior; the Hubble parameter estimates are uniformly high and the slope estimates uniformly low.  Is this a product of triaxiality?  Examining the results obtained when fitting to halos of different degrees of triaxiality at the same orientation (as shown in Figure \ref{plot:figure9}) and those obtained when fits are carried out for fixed slope parameters (as shown in Figure \ref{plot:figure1}), it is clear the answer is no.  Most importantly, the values of $h$ and $\eta$ obtained in the general model do not change significantly with the geometry of the halo -- oblate and prolate halos with the similar visible axis ratios give very similar results.  The visible axis ratio has some impact on parameter estimates due to the changing shape of the posterior probability distribution with respect to the hard wall at $q_p = 1$, but the line-of-sight structure has no impact.  Thus, the offset in parameters cannot be due to the 3D shape of the halo.

Further, it may seem surprising that the most-probable marginalized parameters under a flat prior do not include the true values within their 1$\sigma$ error bars; however, the reliability of the result is underlined by the steady movement of the best-fitting parameter values to the true underlying values as an ever tighter prior is imposed on the profile slope, culminating in the return to the true value when a $\delta$-function prior is imposed and the slope set to isothermal.  Thus, it is not triaxiality, but the large $\eta$-$h$-$b_E$-$q_p$-${\bf \beta}$ degeneracy, that causes the offset in parameter values.  The shapes of the parameter contours tell us that, with a flat prior, a much larger volume of high-likelihood models reside at high values of $h$, low values of $\eta$, low values of $b_E$, and low values of $q_p$ in parameter space.  The effect is strongest for low projected ellipticity halos with values of $q_p$ near 1 due to the hard wall at $q_p=1$, as the contours are pushed proportionally even more to low values of $q_p$ and thus lead to even higher best-fitting values of $h$ and lower values of $\eta$.  Thus the high values of the Hubble parameter are not errors, per se, but the true best estimates of the parameter if the degeneracies in the problem are unbroken.  It is therefore crucial to break the degeneracy in order to obtain meaningful parameter estimates.

\section{Conclusions}
Both cluster and galaxy halos are predicted to exhibit significant triaxiality; in this study we have assessed the impact of neglecting this predicted halo shape on Hubble constant estimation in galaxy strong lensing analyses, and find it is negligible.  Fitting elliptical power law models to triaxial oblate and prolate halos using a flexible MCMC technique shows that, when the density slope of the model matches that of the lensing halo, correct Hubble parameter values are always recovered within the errors without bias, a result that we further explore and explain analytically.  In the case of mis-matched slopes the presence of projected elliptical structure may slightly increase the error in the recovered Hubble parameter values as compared to more spherical cases.  However, prolate and oblate halos of the same projected axis ratio exhibit exactly the same behaviour, in spite of very different line-of-sight geometry, indicating that it is increased projected ellipticity and the resultant shifting of the posterior towards relatively low values of $q_p$ that is significant, not 3D triaxial structure. Thus, we find that halo triaxiality cannot contribute to the inconsistencies between some lensing-derived values of H$_{0}$ and those of the HST Key Project. 
Note that we have considered very triaxial models motivated by (dark matter) N-body simulations, which would represent some of the most triaxial galaxies even when baryonic matter (which plays a crucial role on the scales relevant to strong lensing, see e.g. \citep{deba}) is included.

The Hubble parameter as derived by strong lensing is degenerate not only with the slope of the lensing density profile, but also with the normalisation of the lensing potential, the projected ellipticity of the lens, and the source position.  Accurately constraining the profile slope successfully breaks this significant degeneracy.  However, incorrectly constraining the slope, for example taking an average value across a population as true for a given system, will lead to well-fitting models with incorrect Hubble parameter values, as the degeneracy is broken at the wrong point.  This problem is best resolved by carrying out statistical studies with many lens systems, in which average values of the profile shape are likely to be correct when applied across a population.  Conversely, if the Hubble parameter can be considered well constrained, then strong lensing can be successfully used for profile-type determination (i.e. NFW vs. Isothermal) without requiring the use of complex triaxial models in the fitting process, as our work shows that neither the Hubble parameter or density profile slope estimates are biased by neglected triaxiality.


Past studies have concluded that certain effects can exacerbate the inconsistencies between lensing-derived Hubble parameter values and the HST value; for example, \citet{keet} showed that neglect of environment can cause an \emph{overestimate} in H$_{\rm 0}$ values, the opposite trend of that required to account for the low values observed in many lensing systems. Thus, the remaining factors proposed and still under investigation that may lead to underestimates of the Hubble constant are placed under even greater pressure to explain the discrepancy. One such promising factor is halo substructure, its effects recently demonstrated by \cite{ogurc}; another was highlighted when \cite{dobkeb} showed that, despite findings that in general galaxies strongly tend towards isothermal profiles (\citealt{koop}), group interactions many result in profile steepening that would lead to an  underestimate of H$_{\rm 0}$ should an isothermal slope be assumed.  Out work shows that this effect may be slightly enhanced by highly elliptical lenses, but the additional impact will be small.  Overall, triaxiality cannot explain the difficulties encountered in reconciling disparate lensing results for the Hubble parameter with those of other techniques (e.g. \citealt{kocha}, \citealt{koche}, \citealt{kochd}).  However, the monster degeneracies of the general power law model explored in this paper emphasise one reason why achieving consistent lensing results is extremely difficult without a good external constraint on the density profile slope.

Large samples of multiply imaged sources with measured time delays, deep optical and infrared imaging and spectroscopic data will soon become available. Future space and ground based facilities such as GAIA, LSST, DES, and further down the line, DUNE and the SKA, will greatly enhance our knowledge of the physical and environmental properties of lensing galaxies, may generate lensing samples large enough to allow cleaner studies using only galaxies in isolated environments, and will allow statistical studies of the Hubble parameter in large numbers of lensing systems (e.g., hundreds of systems in DES) in which the strong degeneracies of the general model parameters can be broken, greatly clarifying galaxy lensing's implications for cosmology.

\section*{Acknowledgements}
We thank the Marshall Foundation, the National Science Foundation, and the Cambridge Overseas Trust (VLC), PPARC through a PhD studentship (BMD), and the Royal Society (LJK) for supporting this work. We thank Antony Lewis for very helpful discussions and our anonymous referee for a very useful review process.

\appendix

\section{Parameter estimates and contour shapes as a function of projected axis ratio}\label{sec:appa}
\begin{figure}
\epsfig{file=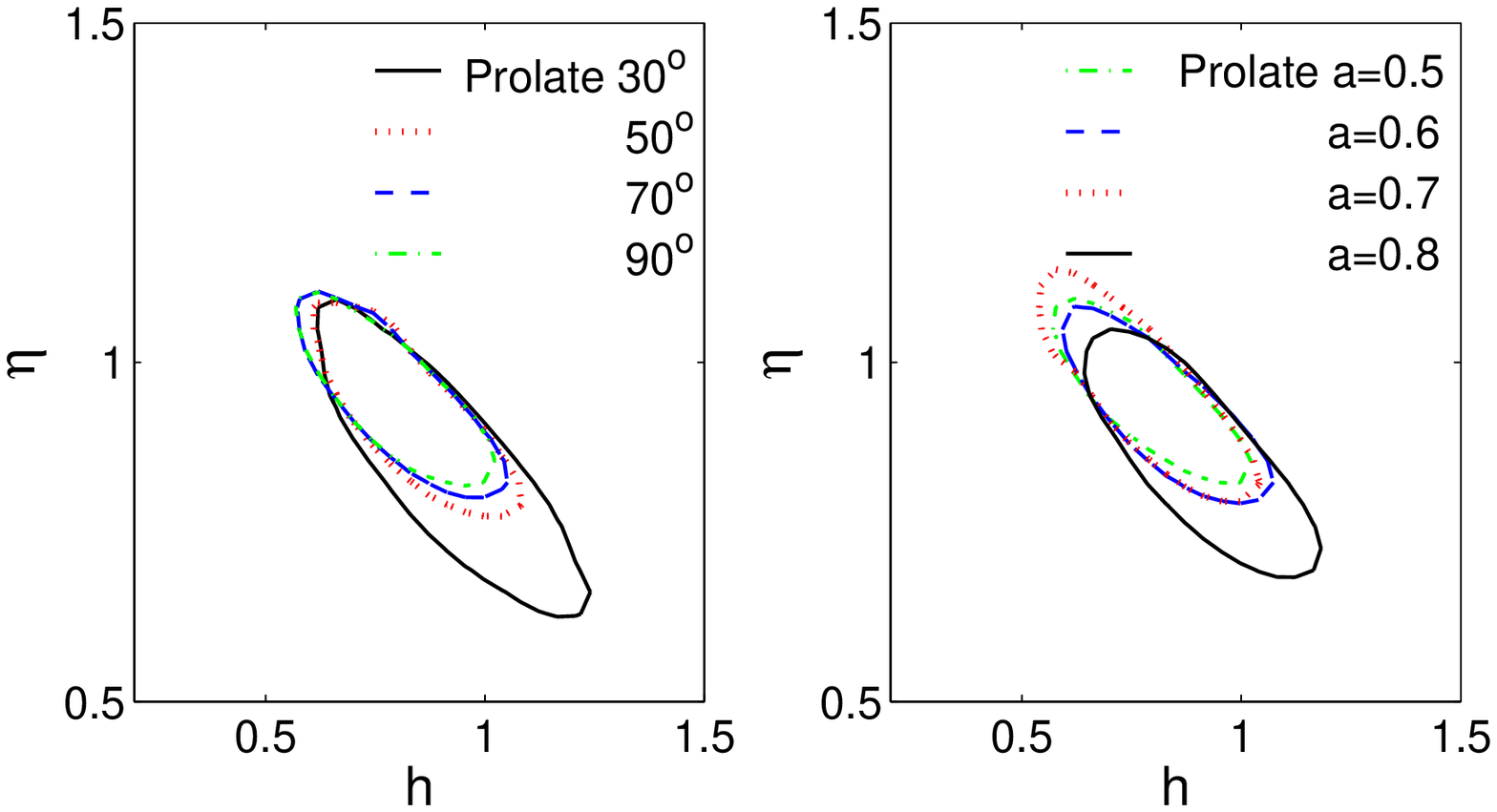,scale=0.45}
\caption{Projections of the posterior probability distribution into the $\eta$-$h$ plane of a power law model fit to an isothermal triaxial prolate halo.  The left-hand panel shows results for lenses of axis ratios ($a=0.6$, $b=0.6$) rotated between orientations of 30$^o$ (low ellipticity, $q_p$ near 1) and 90$^o$ (high ellipticity, lower $q_p$); the right-hand panel shows results for lenses of axis ratios $a=b=\{0.5, 0.6, 0.7, 0.8\}$ all oriented at a high-ellipticity orientation of 80$^o$, with the low $a$ models having the highest projected ellipticities.}
\label{fig:app1}
\end{figure}
\begin{figure}
\epsfig{file=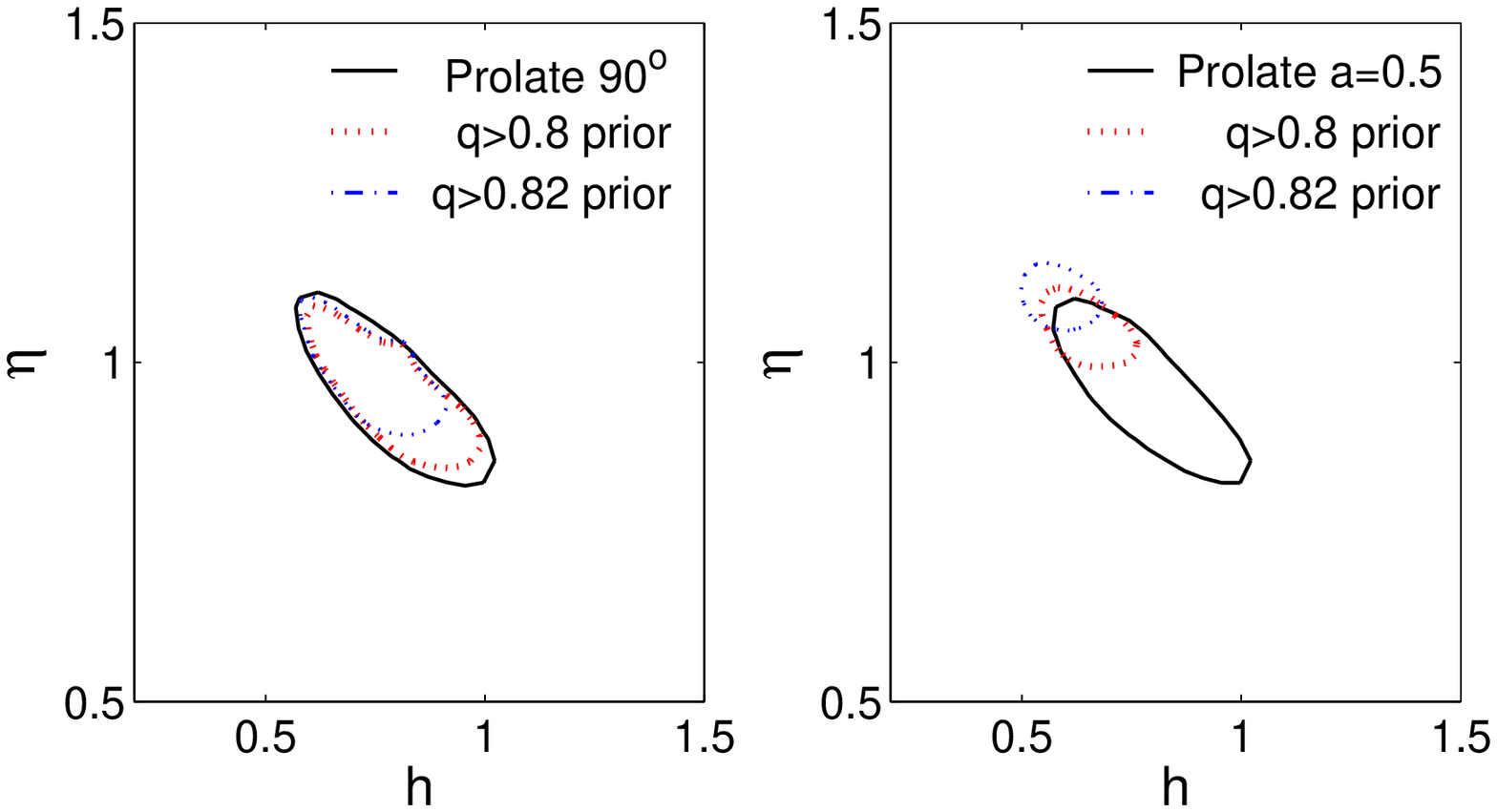,scale=0.45}
\caption{Projections of the posterior probability distribution into the $\eta$-$h$ plane of a power law model fit to an isothermal triaxial prolate halo with three priors on axis ratio $q_p$.  The left-hand panel show results for a lens with axis ratios ($a=0.6$, $b=0.6$) at orientation of 90$^o$ (high ellipticity, lower $q_p$); the right-hand panel shows results for a lens with axis ratios ($a=b=0.5$) oriented at a high-ellipticity orientation of 80$^o$.}
\label{fig:app2}
\end{figure}
When fitting the fully general power-law model, we note a trend in which the most-probable parameter values move closer to the true values with increasing projected ellipticity.  We credit this trend to the presence of a hard wall in the prior on the axis ratio $q_p$ at $q_p=1$, beyond which the equations governing the lensing potential are undefined.  In projected halo geometries with low ellipticities (high $q_p$ values), this wall limits the volume of parameter space available at values of $q_p$ higher than the true value.  This is not an error; it reflects the true shape of the posterior probability distribution.  But it does affect the shape and positioning of the resulting isoprobability surfaces, pushing them towards lower values of $q_p$ relative to the true value, and due to the shape of the $\eta$-$h$-$b_E$-$q_p$-${\bf \beta}$ degeneracy, to higher values of $h$ and ${\bf \beta}$ and lower values of $\eta$.

That this effect depends only on the {\it projected} shape of the potential and not on the 3D geometry is clear when the contours of prolate halos of one geometry projected at different orientation angles are compared to the contours of prolate halos of different underlying geometries (a range of minor axis ratios) all projected at the same orientation.  The $\eta-h$ plane 1$\sigma$ contours for these two cases are shown in Figure \ref{fig:app1}.  In both cases, the projected ellipticity varies from high to low, but the underlying geometry leading to that change is fundamentally different between the two.  However, the exact same trend is visible, with the low ellipticity, high-$q_p$ cases exhibiting larger contours extended toward high values of $h$ and low values of $\eta$.

To further confirm that this behaviour is indeed attributable to the necessary imposition of a hard wall of zero probability at $q_p=1$, we look for the opposite effect in the high ellipticity, low $q_p$ cases, by artificially imposing a hard wall of zero probability at $q_p = \{0.8, 0.82\}$.  If our explanation is correct, the impositions of these priors should produce a similar but opposite effect in the resulting $\eta-h$ contours, shifting them to lower values of $h$ and higher values of $\eta$.  The results for both the constant geometry and constant orientation cases are shown in Figure \ref{fig:app2}, and indeed the contours behave as expected, with the contours moving more and more to higher values of $\eta$ as the prior is tightened.

Thus, the trends seen when lensing through both halos of the same geometries oriented to exhibit different projected ellipticities (Figures \ref{plot:figure6} \& \ref{plot:figure8}) and of different geometries oriented similarly (Figure \ref{plot:figure1}) -- that systems with projected axis ratios close to $q_p= 1$ return parameter values further away from the true underlying value of the system than do their more elliptical counterparts -- is attributable to the real shape of the posterior probability distribution and is a function only of projected ellipticity and independent of 3D structure.

\label{lastpage}
\end{document}